# TRACE: a code for time-reversible astrophysical close encounters

Tiger Lu ,[1]* David M. Hernandez [1] and Hanno Rein [2,3]

[1]*Department of Astronomy, Yale University, 219 Prospect, New Haven, CT 06511, USA*
[2]*Department of Physical and Environmental Sciences, University of Toronto at Scarborough, Toronto, Ontario M1C 1A4, Canada*
[3]*David A. Dunlap Department of Astronomy and Astrophysics, University of Toronto, Toronto, Ontario, M5S 3H4, Canada*

## ABSTRACT

We present TRACE, an almost time-reversible hybrid integrator for the planetary $N$-body problem. Like hybrid symplectic integrators, TRACE can resolve close encounters between particles while retaining many of the accuracy and speed advantages of a fixed time-step symplectic method such as the Wisdom–Holman map. TRACE switches methods time-reversibly during close encounters following the prescription of Hernandez & Dehnen. In this paper we describe the derivation and implementation of TRACE and study its performance for a variety of astrophysical systems. In all our test cases, TRACE is at least as accurate and fast as the hybrid symplectic integrator MERCURIUS. In many cases, TRACE's performance is vastly superior to that of MERCURIUS. In test cases with planet–planet close encounters, TRACE is as accurate as MECURIUS with a 12× speed-up. If close encounters with the central star are considered, TRACE achieves good error performance while MERCURIUS fails to give qualitatively correct results. In ensemble tests of violent scattering systems, TRACE matches the high-accuracy IAS15 while providing a 15× speed-up. In large $N$ systems simulating lunar accretion, TRACE qualitatively gives the same results as IAS15 but at a 41× speed-up. We also discuss some cases such as von Zeipel–Lidov–Kozai cycles where hybrid integrators perform poorly and provide some guidance on which integrator to use for which system. TRACE is freely available within the REBOUND package.

**Key words:** gravitation – methods: numerical – planets and satellites: dynamical evolution and stability.

## 1 INTRODUCTION

The $N$-body problem is one of the most fundamental problems in astronomy. Conceptually, it is a seemingly simple problem: given the initial positions and velocities of $N$ particles, can we predict their state at some arbitrary time in the past or future? In most astronomical contexts, the inter-particle forces are given by Newton's laws of gravitation (Newton 1687). Advancements in our understanding of the $N$-body problem have shed light on topics as varied as the long-term secular behaviour of the Solar system (Laplace 1775; Lagrange 1778), the large-scale structure of the universe (Lemson & Virgo Consortium 2006), and the dynamics of globular clusters (Heggie & Hut 2003), to name but a few.

Despite its conceptual simplicity, solving the $N$-body problem is extremely difficult. The two-body problem is exactly solved (Bernoulli 1775). However, it is well known that for $N \geq 3$ the $N$-body problem admits no practical general analytic solution,[1] with solutions either only valid in the limit of certain simplifying assumptions (Poincaré 1890) or slow to the point of being completely infeasible in practice (Sundman 1913; Qiu-Dong 1990). With these constraints of both accuracy and computation time, to study the $N$-body problem we must turn to numerical methods of approximation.

Of particular interest to astronomers is the planetary $N$-body problem, which is characterized by a dominant central 'star' orbited by many smaller 'planets'. Wisdom & Holman (1991)[2] developed an efficient, accurate and widely used integrator for the planetary $N$-body problem by treating the effects of other planets in the system as perturbations to the dominant Keplerian motion. Improvements on this 'Wisdom–Holman' method over the years are described in Saha & Tremaine (1992, 1994); Wisdom, Holman & Touma (1996); Laskar & Robutel (2001); Hernandez & Bertschinger (2015); Rein & Tamayo (2015); Hernandez (2016); Wisdom (2018); Rein, Tamayo & Brown (2019b); Jahaveri, Rein & Tamayo (2023). The Wisdom–Holman method is an example of a symplectic method, from which many of its desirable characteristics can be attributed to. Symplectic integrators solve Hamiltonian systems, and are hugely advantageous because they exactly conserve phase space volumes and Poincaré invariants (Yoshida 1993; Hairer et al. 2006). Due to these constraints, they boast impressive energy error performance over millions of dynamical time-scales, whereas conventional integrators may exhibit significant failures after only a few. Given that the dynamics of most astrophysical systems are governed by Hamiltonians (as far as gravity is concerned), symplectic integrators are ideal for their study. The Wisdom–Holman scheme allowed for feasible computation of the evolution of planetary systems on Gyr time-scales, and its speed and efficiency have made insights into computationally demanding topics such as the stability of planetary systems (Holman & Wisdom 1993; Holman & Wiegert 1999) and the precise orbital and obliquity evolution of solar system planets (Touma & Wisdom 1993; Laskar et al. 2004, 2011) possible.

---

* E-mail: tiger.lu@yale.edu

[1]Newton implies that the $N$-body problem is in general unsolvable in his original manuscript.

[2]Kinoshita, Yoshida & Nakai (1991) developed a similar integrator independently.

Wisdom–Holman integrators have become mainstays in celestial dynamics, but this is not to say they are without their drawbacks. One such shortcoming is inflexibility: usually symplectic methods use a constant time-step which cannot be adapted if relevant time-scales in the problem change. Focusing on the Wisdom–Holman method in particular, it fails when the underlying assumption of a dominant Keplerian orbit is challenged. This occurs primarily when there is a close encounter between two pairs of bodies and interparticle forces dominate instead. In practice, this means Wisdom–Holman integrators are only effective for problems where all planets maintain stable Keplerian orbits for the duration of the simulation. There are a few ways to circumvent this issue. We will focus on the method of hybrid symplectic integrators. Schemes such as MERCURY/MERCURIUS (Chambers 1999; Rein et al. 2019a) and modified SYMBA (Duncan, Levison & Lee 1998; Levison & Duncan 2000) are able to achieve acceptable levels of accuracy while retaining many of the long-term error and speed benefits of the Wisdom–Holman map in uses cases when the traditional map fails. This is achieved by using maps utilizing more flexible conventional (but non-symplectic) integrators such as Bulirsch–Stoer (Press et al. 2002) or IAS15 (Rein & Spiegel 2015) upon close encounters, switching between integration methods based on some pre-determined switching function. Hybrid symplectic integrators have allowed for study of topics such as planetary/lunar accretion (Canup 2004; Raymond, Quinn & Lunine 2006), the dynamical history of our solar system (Gomes et al. 2005; Morbidelli 2005; Tsiganis et al. 2005), and the imprint of instabilities on the demographics of exoplanetary systems (Chatterjee et al. 2008; Ford & Rasio 2008; Lissauer et al. 2011), all topics involving regimes where the Wisdom–Holman method breaks down but are simultaneously too computationally demanding for conventional integrators. The hybrid integrators listed are symplectic, which come with advantages and disadvantages. Maintaining symplecticity ensures good long-term error performance, but also carries with it much of the inflexibility described before. This primarily manifests in the choice of switching function, which in practice can often be restrictive or cumbersome.

In this work, we take an alternative approach to constructing a hybrid integrator. Our integrator is not symplectic, but instead is time-reversible. While the breadth of study and literature on time-reversible integration methods is small in comparison to that of symplectic methods (Hut, Makino & McMillan 1995; Hairer et al. 2006; Hairer, McLachlan & Skeel 2009; Dehnen 2017; Hernandez & Bertschinger 2018; Boekholt, Vaillant & Correia 2023), in principle an exactly time-reversible scheme shares many of the long-term beneficial error properties as a symplectic integrator. While many time-reversible schemes have proven computationally infeasible for practical purposes, in the recent work of Hernandez & Dehnen (2023) a simple time-reversible algorithm was developed with comparable error and computational performance to symplectic methods.

Based on the ideas of Hernandez & Dehnen (2023) we present TRACE, an almost time-reversible hybrid integrator for the planetary *N*-body problem. TRACE is not exactly symplectic (see Appendix C for an analysis). The TRACE algorithm and switching scheme is conceptually simple, flexible, and easy to modify. It is capable of accurately integrating close encounters between any pair of bodies in the planetary *N*-body problem, including the central star. We have tested TRACE on a variety of realistic astrophysical systems. We have further developed switching functions beyond those discussed in Her-nandez & Dehnen (2023), performed statistical tests on ensembles of chaotic scattering systems, and tested the performance limits of our code on large *N* systems. In all cases, TRACE matches or exceeds the accuracy of previous hybrid integrators such as MERCURIUS, and admits speed-ups of up to 13×. TRACE is publicly available in the REBOUND *N*-body package (Rein & Liu 2012). The structure for this paper is as follows. In Section 2, we provide background on the construction of integrators for the planetary *N*-body problem. In Section 3, we discuss the current hybrid integration techniques currently available and derive the TRACE equations of motion. In Section 4, we discuss the TRACE switching scheme. In Section 5, evaluate TRACE's performance on realistic astrophysical systems one might encounter including highly eccentric orbits, planet−planet scattering and planetesimal accretion. In Section 6, we discuss potential improvements to the TRACE algorithm. In Section 7, we draw our conclusions and provide specific guidelines for when TRACE should be used over other integrators.

## 2 EQUATIONS OF MOTION OF THE PLANETARY *N*-BODY PROBLEM

In this section, we introduce the equations of motion used for the planetary *N*-body problem, and review the construction and benefits of the Wisdom–Holman map.

### 2.1 Constructing Hamiltonian maps

We will first provide a brief overview of constructing maps for conservative Hamiltonian systems in general. Consider one system governed by the Hamiltonian $\mathcal{H}$. Denote the state of the system in canonical coordinates by the vector $z = (q, p)$. Hamilton's equations dictate the time-evolution of $z$,

$$\frac{dz}{dt} = \{z, \mathcal{H}\}, \tag{1}$$

Where $t$ is time. The *Lie operator* $\hat{\mathcal{H}}$ is defined as $\hat{\mathcal{H}}z \equiv \{z, \mathcal{H}\}$. We can thus rewrite equation (1),

$$\frac{dz}{dt} = \hat{\mathcal{H}}z, \tag{2}$$

This differential equation admits the solution,

$$z(t + h) = e^{h\hat{\mathcal{H}}}z(t). \tag{3}$$

Here, $h$ is referred to as the time-step and $e^{h\hat{\mathcal{H}}}$ is defined as the *propagator* or *map*.[3] While this is indeed an exact solution for the problem, in many cases this is difficult to solve and impractical. One useful path forward is operator splitting: in many Hamiltonian systems, we may decompose $\mathcal{H}$ into the sum of sub-Hamiltonians, each corresponding to some component of the motion. A simple scheme is to split the potential and kinetic components,

$$\mathcal{H} = T(p) + V(q). \tag{4}$$

In practice, these sub-Hamiltonians are often significantly simpler to solve, and in many cases admit analytic solutions where the full Hamiltonian does not. The idea behind operator splitting is that the true equation of motion may be approximated by first evolving the system under $T$, and then $V$. Maps may be constructed through different splittings and applications of the individual propagators over various time-steps. Splitting schemes are not exact solutions, but are often the only viable way to study such systems. The error in a splitting scheme can be analysed via the Baker–Campbell–Hausdorff (BCH) formula (Campbell [1897](#); Baker [1905](#); Hausdorff [1906](#); Hairer et al. [2006](#)). For propagators $\hat{A}$, $\hat{B}$, and $\hat{C}$ satisfying $\hat{C} = \hat{A} + \hat{B}$, the local error in the canonical coordinates over one step of the splitting scheme can be expressed as

$$
\begin{aligned}
\text{Error} &= \left( e^{h\hat{C}} - e^{h\hat{A}} e^{h\hat{B}} \right) z \\
&= \frac{h^2}{2} [\hat{A}, \hat{B}] z + O(h^3),
\end{aligned}
\tag{5}
$$

where $[\hat{A}, \hat{B}] \equiv \hat{A}\hat{B} - \hat{B}\hat{A}$ is the commutator. Higher order terms in $h$ depend on a series of nested commutators, so the error in a splitting scheme arises from pairs of propagators not commuting. For instance, the well-known leapfrog method is given $e^{h\hat{\mathcal{H}}} \approx e^{\frac{h}{2}\hat{T}} e^{h\hat{V}} e^{\frac{h}{2}\hat{T}}$. By using the convenient symmetric form of the BCH formula, we can calculate the error in one time-step of leapfrog,

$$
\begin{aligned}
\text{Leapfrog Error} &= \left( e^{h\hat{\mathcal{H}}} - e^{\frac{h}{2}\hat{T}} e^{h\hat{V}} e^{\frac{h}{2}\hat{T}} \right) z \\
&= \frac{h^3}{24} \left( [\hat{V}, [\hat{V}, \hat{T}]] - 2[\hat{T}, [\hat{T}, \hat{V}]] \right) z + O(h^5).
\end{aligned}
\tag{6}
$$

---

[3]Explicitly, the propagator is defined by the Taylor series $e^{h\hat{\mathcal{H}}} \equiv \sum_{n=0}^{\infty} \frac{\hat{\mathcal{H}}^n}{n!}$.

Note that symmetry ensures no even powers of $h$ survive in this expansion. Errors that depend higher power of $h$ are preferable as the time-step is generally small.

## 2.2 The Wisdom–Holman map

The planetary $N$-body problem considers a system of $N$ planets with a dominant central mass. The central mass is denoted with subscript 0, and the other planets 1, 2, ..., $N$. The well-known Hamiltonian of the system may be written

$$
\mathcal{H} = \sum_{0 \le i} \frac{p_i^2}{2m_i} - G \sum_{0 \le i < j} \frac{m_i m_j}{|\boldsymbol{q}_i - \boldsymbol{q}_j|},
\tag{7}
$$

with $\boldsymbol{q}$, $\boldsymbol{p}$ the canonical coordinates/momenta and $\boldsymbol{m}$ the masses. We make an important note at this point: this is the **only** Hamiltonian considered in this work. We will rewrite this Hamiltonian many times for convenience, but all will be exactly equal to equation (7).

Of course, the full $N$-body Hamiltonian is very difficult to numerically solve. The brilliance of the Wisdom–Holman map is in its clever splitting of the Hamiltonian into a dominant and a much smaller part which may be considered a perturbation. Specifically, the gravity of the Sun is considered the dominant part and the influence of the other planets in the system are considered perturbations. In this sense, the Wisdom–Holman map approximates the planetary $N$-body problem into $N$ individual Kepler problems, one for each planet. It is clear why this is effective upon inspection of the BCH formula. If the Hamiltonian is split into two parts $\mathcal{H} = \mathcal{H}_1 + \mathcal{H}_2$ such that $\mathcal{H}_2 = \epsilon \mathcal{H}_1$ with $\epsilon \ll 1$, then the local error over one time-step will scale as $O(\epsilon h^3)$, in comparison with $O(h^3)$ as in standard leapfrog (Wisdom & Holman [1991](#); Tremaine [2023](#)). This allows Wisdom–Holman integrators to take comparatively large time-steps while maintaining small errors. The resulting speed and accuracy has allowed for long-term integrations of planetary systems on time-scales comparable to the age of the solar system.

We will make use of democratic heliocentric coordinates (DHC) $\boldsymbol{Q}_i$ and momenta $\boldsymbol{P}_i$. For more in-depth discussion of this coordinate system, see Duncan et al. ([1998](#)), Hernandez & Dehnen ([2017](#)), Rein & Tamayo ([2019](#)), as well as Appendix A. Note that the Wisdom–Holman map was originally derived in Jacobi coordinates instead of DHC – we use the DHC coordinate system because it can

effectively deal with orbit crossings. Our state vector is thus defined:

$$
z_i \equiv \left( \boldsymbol{Q}_i, \boldsymbol{P}_i \right).
\tag{8}
$$

The advantage of DHC is that it allows us to rewrite equation (7) as a sum of four terms, each with a clear physical interpretation,

$$
\mathcal{H} = \underbrace{\frac{P_0^2}{2m_{\text{tot}}}}_{\mathcal{H}_0} + \underbrace{\frac{1}{2m_0} \left( \sum_{i \ne 0} P_i \right)^2}_{\mathcal{H}_J} - \underbrace{\sum_{0 < i < j} \frac{Gm_i m_j}{Q_{ij}}}_{\mathcal{H}_I} \\
+ \underbrace{\sum_{i > 0} \left( \frac{P_i^2}{2m_i} - \frac{Gm_0 m_i}{Q_i} \right)}_{\mathcal{H}_K}.
\tag{9}
$$

We use underbraces to mark the sub-Hamiltonians for clarity. Here, $\mathcal{H}_0$ describes the motion of the centre of mass, $\mathcal{H}_J$ is called the jump term and describes the barycentric motion of the star, $\mathcal{H}_I$ describes planet–planet interactions, $\mathcal{H}_K$ the pure Keplerian motion of the planets around the central body. $m_{\text{tot}} = \sum_{i=0}^{N} m_i$ is the total mass of the system), and $\boldsymbol{Q}_{ij} \equiv \boldsymbol{Q}_i - \boldsymbol{Q}_j$. In DHC, the splitting of the Wisdom–Holman map is given by the following composition, which uses the splitting in equation (9):

$$
M = e^{\frac{h}{2}\hat{\mathcal{H}}_I} e^{\frac{h}{2}\hat{\mathcal{H}}_J} e^{h\hat{\mathcal{H}}_0} e^{h\hat{\mathcal{H}}_K} e^{\frac{h}{2}\hat{\mathcal{H}}_J} e^{\frac{h}{2}\hat{\mathcal{H}}_I}.
\tag{10}
$$

Each of the sub-Hamiltonians $\mathcal{H}_0$, $\mathcal{H}_J$, $\mathcal{H}_I$, and $\mathcal{H}_K$ may individually be solved analytically. The equations of motion governed by $\mathcal{H}_0$, $\mathcal{H}_J$ and $\mathcal{H}_I$ are trivially solved, while $\mathcal{H}_K$ corresponds to Kepler's equations which can be solved with a Kepler solver (Danby [1992](#)).

# 3 HYBRID INTEGRATORS

There are a variety of relevant and interesting situations in many astrophysical systems where this assumptions underlying the Wisdom–Holman integrator break down. As mentioned previously, more conventional integrators are better equipped to handle these failure cases, but lose the long-term error benefits of Wisdom–Holman.

Ideally, we would like to use the Wisdom–Holman scheme when possible to leverage its considerable speed advantages, and use a more conventional, flexible integrator when the assumptions inherent to the Wisdom–Holman map break down in the interest of accuracy. This is the idea behind hybrid integrators. In this section, we will enumerate various situations where the Wisdom–Holman map fails, and discuss the existing solutions. At the end, we will introduce our map TRACE, which is capable of effectively handling all such pitfalls.

## 3.1 Planet–planet close encounters

If two planets undergo a close encounter, $\mathcal{H}_I$ will dominate over $\mathcal{H}_K$. The Wisdom–Holman scheme breaks down here. A solution was proposed by Chambers ([1999](#)) in their code MERCURY by smoothly moving the interaction term to the Kepler term. This ensures that $\mathcal{H}_K$ is always the dominant term in the Hamiltonian. A modified version of this scheme is implemented in REBOUND in the form of the hybrid symplectic integrator MERCURIUS (Rein et al. [2019a](#)). While $\mathcal{H}_K$ no longer exactly corresponds to Kepler's equation and cannot be analytically solved, it is possible to approximate accurately and efficiently with a more conventional integration techniques. MERCURY uses a Bulirsch–Stoer scheme while MERCURIUS uses IAS15, an adaptive-time-step 15th-order non-symplectic integrator that serves as the default integrator in REBOUND. Explicitly, the MERCURIUS map is obtained by splitting equation (7), for use in map (10), in the following way,

$$
\mathcal{H} = \underbrace{\frac{P_0^2}{2m_{\mathrm{tot}}}}_{\mathcal{H}_0} + \underbrace{\frac{1}{2m_0}\left(\sum_{i\neq 0} P_i\right)^2}_{\mathcal{H}_J}
$$

$$
- \underbrace{\sum_{0<i<j} \frac{Gm_im_j}{Q_{ij}}\left[1 - K(\boldsymbol{Q}_{ij})\right]}_{\mathcal{H}_I}
$$

$$
+ \underbrace{\sum_{i>0}\left(\frac{P_i^2}{2m_i} - \frac{Gm_0m_i}{Q_i}\right) - \sum_{0<i<j}\frac{Gm_im_j}{Q_{ij}}K(\boldsymbol{Q}_{ij})}_{\mathcal{H}_K}. \tag{11}
$$

Note that this is exactly equal to equation (9) as the terms with $K(\boldsymbol{Q}_{ij})$ cancel, but we have redefined the sub-Hamiltonians that affect the splitting scheme equation (10). Here, the centre of mass and jump terms remain the same as those of the Wisdom–Holman map. Meanwhile, the Kepler and interaction terms are now modulated by the switching function $K(\boldsymbol{Q}_{ij})$, a mathematically smooth scalar function that is purely a function of the pairwise distance between the two bodies in question and takes values $\in [0, 1]$. MERCURIUS offers several built-in switching functions, but all smoothly switch from $K = 1$ at close encounters to $K = 0$ very far from an encounter. Note that for $K = 0$, the standard Wisdom–Holman map is recovered.

## 3.2 Pericenter approach

The Wisdom–Holman map in DHC encounters issues for massive particles on orbits with very close pericentric distances.[4] This is because $\mathcal{H}_K$ does not exactly represent a Keplerian orbit, since it incorporates a non-physical central gravitating mass – $\mathcal{H}_J$ must be incorporated as well to correct. Hence, when $\mathcal{H}_J$ becomes very large during close pericenter approaches the Wisdom–Holman method fails as well (Duncan et al. 1998; Rauch & Holman 1999). In principle, it is possible to avoid this issue by resolving the pericenter with a small enough time-step (Wisdom 2015). However, since Wisdom–Holman uses a fixed time-step this worst-case time-step must be applied to the entire problem that comes at a significant computational cost. Neither MERCURY nor MERCURIUS allow for close pericenter approaches.

There are two approaches that one can take to resolve this issue. Levison & Duncan (2000) propose a solution in which entails smoothly moving terms from the jump term to the Kepler term upon a close encounter with the central body. Explicitly, their map is obtained by splitting (7), for use in map (10), in the following way,

$$
\mathcal{H} = \underbrace{\frac{P_0^2}{2m_{\mathrm{tot}}}}_{\mathcal{H}_0} - \underbrace{\sum_{0<i<j}\frac{Gm_im_j}{Q_{ij}}}_{\mathcal{H}_I}
$$

$$
+ \underbrace{\frac{1}{2m_0}\left(\sum_{i\neq 0} P_i\right)^2 (1 - F(\boldsymbol{Q}))}_{\mathcal{H}_J}
$$

$$
+ \underbrace{\sum_{i>0}\left(\frac{P_i^2}{2m_i} - \frac{Gm_0m_i}{Q_i}\right) + \frac{1}{2m_0}\left(\sum_{i\neq 0} P_i\right)^2 F(\boldsymbol{Q})}_{\mathcal{H}_K}. \tag{12}
$$

[4]In Jacobi coordinates, Wisdom–Holman can integrate arbitrarily eccentric orbits as long as the interaction term is 0. For more discussion see (Duncan et al. 1998).

Compared to the classic Wisdom–Holman map the centre of mass and interaction terms do not change. The jump and Kepler terms are modulated by $F$, again a mathematically smooth function taking values $\in [0, 1]$ of all particles' heliocentric distances, with $F = 1$ very close to the central body and $F = 0$ very far from it. Hernandez & Dehnen (2023) expand on this method by using a discrete binary switching function for $F$. Again, note that the Wisdom–Holman map is recovered in the case of $F = 0$. Note also that $F(\boldsymbol{Q})$ is a function of all particles' pericenter distance, in contrast to $K(\boldsymbol{Q}_{ij})$ in MERCURY/MERCURIUS. This is because when any particle undergoes a pericenter passage, the jump term must be shifted to the Kepler term. As the jump term is a function of all $P_i$ in the system, this means the Kepler term of the particle is now coupled to every other particle in the system – see equation (22) – and cannot be independently integrated. Since we only need to integrate $\mathcal{H}_K$ with the conventional integrator and stay within the DHC coordinate system, we denote this as the PARTIAL PERI approach.

We have found in our testing (see Section 5) that this solution, although not failing, achieves less than desirable results for some cases such as massive bodies on very eccentric orbits due to numerical instabilities. Therefore, we present and find good success with an alternative approach. In this approach, when a close approach with the central star is detected we abandon DHC coordinates entirely and perform our integration in the inertial frame. Explicitly, if $F(\boldsymbol{Q}) = 0$ we integrate the standard Wisdom–Holman map in DHC coordinates (equation 9). If $F(\boldsymbol{Q}) = 1$, we convert our system back to the inertial frame and integrate equation (1) with a conventional integrator. This approach completely sidesteps all issues with the DHC splitting for close pericenter approaches. We will show below that this approach is slightly slower than the Levison & Duncan (2000) and Hernandez & Dehnen (2023) method, since we are including all interaction terms in the more complex $\mathcal{H}$, but well worth the trade-off in accuracy. Since we are completely switching integration schemes, we denote this as the FULL PERI approach.

## 3.3 The trace maps

We combine the above concepts from Chambers (1999), Levison & Duncan (2000), Hernandez & Dehnen (2023) as well as our new FULL PERI switching criteria to derive the TRACE map. TRACE can work in two regimes depending on the state of the system: in DHC coordinates, and in inertial Cartesian coordinates (hereafter, simply referred to as 'inertial'). The evolution in each of these regimes is described by the following splittings of (7), for use in map (10), respectively:

$$
\mathcal{H} = \underbrace{\frac{P_0^2}{2m_{\mathrm{tot}}}}_{\mathcal{H}_0} + \underbrace{\frac{1}{2m_0}\left(\sum_{i\neq 0} P_i\right)^2\left[1 - C|_{t_n}\right]}_{\mathcal{H}_J} - \underbrace{\sum_{0<i<j}\frac{Gm_im_j}{Q_{ij}}\left[1 - \mathcal{K}_{ij}|_{t_n}\right]}_{\mathcal{H}_I}
$$

$$
+ \underbrace{\sum_{i>0}\left(\frac{P_i^2}{2m_i} - \frac{Gm_0m_i}{Q_i}\right) - \sum_{0<i<j}\frac{Gm_im_j}{Q_{ij}}\mathcal{K}_{ij}|_{t_n} + \frac{1}{2m_0}\left(\sum_{i\neq 0} P_i\right)^2 C|_{t_n}}_{\mathcal{H}_K}, \tag{13}
$$

and,

$$
\mathcal{H} = \underbrace{\sum_i \frac{p_i^2}{2m_i} - G\sum_{0\leq i<j}\frac{m_im_j}{|\boldsymbol{q}_i - \boldsymbol{q}_j|}}_{\mathcal{H}_K}, \tag{14}
$$

where $\mathcal{H}_0 = \mathcal{H}_J = \mathcal{H}_I = 0$. Here, $\mathcal{K}$ and $C$ are the splitting functions for planet–planet and planet–star encounters, respectively. There is a major difference in writing equation (13) compared with equations

(11) and (12): the functions $C$ and $\mathcal{K}_{ij}$ are no longer smooth functions of the phase space, but are instead modified Heaviside functions, in which their value is allowed to be 0 and 1 only and updated immediately before a time-step only: we have denoted the Heaviside functions with $C\big|_{t_n}$ and $\mathcal{K}_{ij}\big|_{t_n}$. $t_n$ indicates the $n$th discrete time sampled by the integrator, or the simulation time immediately pre-time-step. To reduce clutter, we will drop the $t_n$ superscript for the remainder of the paper with the understanding that all $\mathcal{K}_{ij}$ and $C$ are evaluated only at $t = t_n$.

Hernandez & Dehnen ([2023](#)) demonstrated that velocity-dependent switching functions are viable for such time-reversible, but not symplectic codes, and in this work we show that these switching functions can actually depend on higher derivatives of position. Hence, we will add a dependence on $Q^{(x)}$ to show that arbitrary derivatives of $Q$ may be accounted for in both switching functions. In this section, for brevity we will use the shorthands $\mathcal{K}\left(Q_{ij}^{(x)}\right) \equiv \mathcal{K}_{ij}$ and $C\left(Q^{(x)}\right) \equiv C$. $\mathcal{K}_{ij} = 1$ if there is a planet–planet close encounter between planets $i$ and $j$, and $\mathcal{K}_{ij} = 0$ otherwise. Similarly, $C = 1$ if there is any close encounter with the central body, and $C = 0$ otherwise.

We work in DHC coordinates whenever there is no close encounter with the central body $C = 0$, or if the PARTIAL PERI prescription is used. We only work in the inertial frame if there is both a close pericenter approach ($C = 1$) and we are using the FULL PERI prescription.

Note that our splitting functions are discrete rather than smooth. Discrete switching functions were analysed in Hernandez ([2019](#)), and found to generally perform inaccurately in the long term when compared to continuous, smooth switching functions (as seen in MERCURIUS and SYMBA). However, we will see that with the reversible switching scheme of Hernandez & Dehnen ([2023](#)) that we have implemented, the discrete switching function has comparable error performance to the continuous case, in contrast to the results of Hernandez ([2019](#)). This allows us to leverage the conceptually simpler discrete switching function. We describe our switching algorithm in depth in the following section.

We may use Hamilton's equations to derive the equations of motion associated with the components $\mathcal{H}_0, \mathcal{H}_J, \mathcal{H}_I, \mathcal{H}_K$ that make up $\mathcal{H}$. For the CoM step,

$$\dot{\boldsymbol{Q}}_0 = \frac{\boldsymbol{P}_0}{m_{\text{tot}}}, \tag{15}$$

$$\dot{\boldsymbol{V}}_0 = 0. \tag{16}$$

For the jump step,

$$\dot{\boldsymbol{Q}}_i = \frac{\partial \mathcal{H}_J}{\partial \boldsymbol{P}_i} = \frac{1}{m_0}\left(\sum_{k>0} \boldsymbol{P}_k\right)[1 - C], \tag{17}$$

$$\dot{\boldsymbol{V}}_i = -\frac{1}{m_i}\frac{\partial \mathcal{H}_J}{\partial \boldsymbol{Q}_i} = 0. \tag{18}$$

For the interaction step,

$$\dot{\boldsymbol{Q}}_i = 0, \tag{19}$$

$$\dot{\boldsymbol{V}}_i = -G\sum_{j\neq i, j\neq 0}\frac{m_j}{Q_{ij}^3}\boldsymbol{Q}_{ij}\left[1 - \mathcal{K}_{ij}\right]. \tag{20}$$

And finally for the Kepler step $\mathcal{H}_K$,

$$\dot{\boldsymbol{Q}}_i = \frac{1}{m_0}\left(\sum_{k>0}\boldsymbol{P}_k\right)C + \boldsymbol{V}_i, \tag{21}$$

$$\dot{\boldsymbol{V}}_i = -\frac{Gm_0}{Q_i^3}\boldsymbol{Q}_i - \sum_{j\neq i, j\neq 0}\left(\frac{Gm_j}{Q_{ij}^3}\boldsymbol{Q}_{ij}\right)\mathcal{K}_{ij}. \tag{22}$$

Here, $\boldsymbol{V}_i \equiv \boldsymbol{P}_i/m_i$ are the heliocentric velocities. For $C = \forall(i, j)\mathcal{K}_{ij} = 0$, all components of $\mathcal{H}$ admit analytic solutions. $\mathcal{H}_K$ is the only non-trivial equation of motion, and is solved with the fast Kepler solver used by WHFAST. In the case of $C = 1$ or $\mathcal{K} = 1$, $\mathcal{H}_K$ becomes non-integrable and is solved with the BS implementation in REBOUND, which was first implemented in Lu et al. ([2023](#)). In the $0 < C < 1$ and $0 < \mathcal{K} < 1$ regimes $\mathcal{H}_J$ and $\mathcal{H}_I$ also become non-integrable, respectively. The discrete switching function completely avoids this regime – in this scheme only $\mathcal{H}_K$ will be non-integrable and computationally expensive to solve. The equations of motion associated with $\mathcal{H}_{\text{Inertial}}$ are significantly more complex, and are always expensive to solve. We always require the use of a conventional integrator such as BS or IAS15 to solve the equations in the inertial frame.

## 4 THE TRACE CODE

### 4.1 Switching scheme

As previously mentioned, the computational benefits of the discrete switching function typically come with the trade-off of poor error performance. The time-reversible algorithm by Hernandez & Dehnen ([2023](#)) sidesteps this issue by changing integrators reversibly upon a close encounter, and achieves better error performance with reduced computational cost and conceptual simplicity. In this section, we provide a brief summary of the algorithm, and describe our specific switching functions.

Fig. [1](#) schematically walks through the switching algorithm. Let us define each of these terms for the TRACE map specifically:

(i) $M_1$ is the map used when there is no pericenter approach. In other words, $C = 0$, but we may have some pairs of $\mathcal{K}_{ij} = 1$. Explicitly, this map is obtained by splitting ([7](#)), for use in map ([10](#)), in the following way:

$$\mathcal{H} = \underbrace{\frac{P_0^2}{2m_{\text{tot}}}}_{\mathcal{H}_0} + \underbrace{\frac{1}{2m_0}\left(\sum_{i\neq 0}P_i\right)^2}_{\mathcal{H}_J} - \underbrace{\sum_{0<i<j}\frac{Gm_im_j}{Q_{ij}}\left[1 - \mathcal{K}_{ij}\right]}_{\mathcal{H}_I}$$
$$+ \underbrace{\sum_{i>0}\left(\frac{P_i^2}{2m_i} - \frac{Gm_0m_i}{Q_i}\right) - \sum_{0<i<j}\frac{Gm_im_j}{Q_{ij}}\mathcal{K}_{ij}}_{\mathcal{H}_K}. \tag{23}$$

The jump step is applied to all particles, while the interaction step is applied to particle pairs for which $\mathcal{K}_{ij} = 0$. For particles, $i$ that are not undergoing any planet–planet close encounters ($\mathcal{K}_{ij} = 0$ for all $j$) the Kepler step is solved with the WHFAST Kepler solver. Otherwise, BS is used.

(ii) $M_2$ is the map used when there is a close pericenter approach ($C = 1$), and we are using the PARTIAL PERI prescription. Explicitly, this map is obtained by splitting ([7](#)), for use in map ([10](#)), in the following way,

$$\mathcal{H} = \underbrace{\frac{P_0^2}{2m_{\text{tot}}}}_{\mathcal{H}_0} - \underbrace{\sum_{0<i<j}\frac{Gm_im_j}{Q_{ij}}\left[1 - \mathcal{K}_{ij}\right]}_{\mathcal{H}_I}$$
$$+ \underbrace{\sum_{i>0}\left(\frac{P_i^2}{2m_i} - \frac{Gm_0m_i}{Q_i}\right) - \sum_{0<i<j}\frac{Gm_im_j}{Q_{ij}}\mathcal{K}_{ij} + \frac{1}{2m_0}\left(\sum_{i\neq 0}P_i\right)^2}_{\mathcal{H}_K}. \tag{24}$$

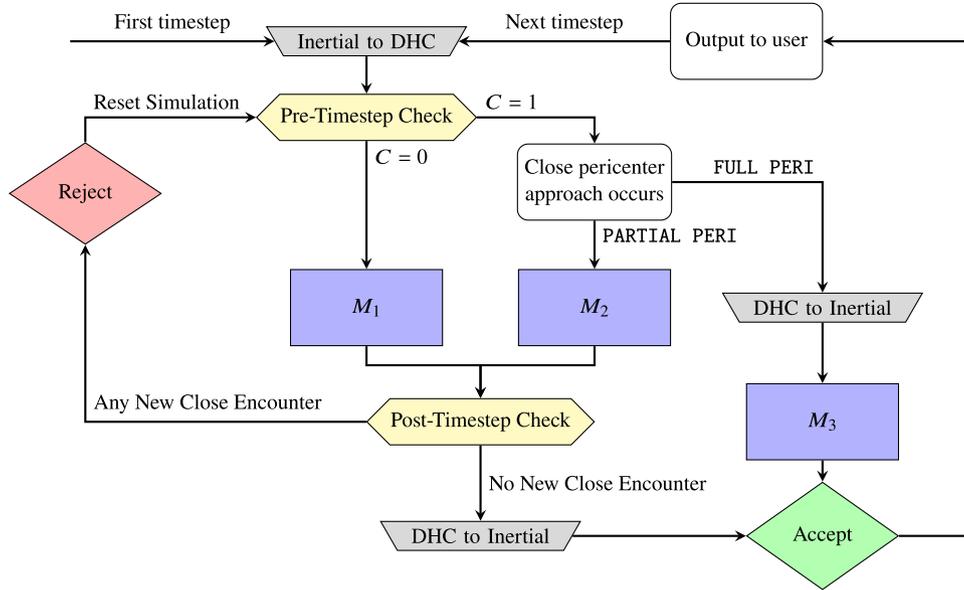

**Figure 1.** Flowchart of the TRACE algorithm. Grey trapezoids correspond to shifting reference frames. Yellow hexagons correspond to checking the planet–planet and pericenter close encounter conditions. The red diamond corresponds to step rejections that result in resetting the simulation to the pre-time-step conditions. The green diamond represents a step acceptance. Blue rectangles correspond to advancements of the simulation with the various maps $M_1$, $M_2$, $M_3$. A *New Close Encounter* is defined as either any $\mathcal{K}_{ij}$ or $C_i$ which was previously evaluated as 0 now evaluated as 1.

$\mathcal{H}_J = 0$, while the interaction step is applied to particle pairs for which $\mathcal{K}_{ij} = 0$ as previously. The Kepler step for all particles is integrated using BS.

(iii) $M_3$ is the map used when there is a close pericenter approach ($C = 1$), and we are using the FULL PERI prescription. Explicitly,

$$\mathcal{H} = \underbrace{\sum_i \frac{p_i^2}{2m_i} - G \sum_{0 \le i < j} \frac{m_i m_j}{|\boldsymbol{q}_i - \boldsymbol{q}_j|}}_{\mathcal{H}_K}. \tag{25}$$

The DHC coordinates are abandoned here, and we simply integrate as in equation (14). TRACE offers two options for this map: BS and IAS15. To differentiate these two options we denote them FULL BS and FULL IAS15.

We will now describe the switching algorithm in detail. By default, TRACE uses the FULL BS pericenter prescription.

(i) The system is first converted from inertial coordinates to DHC. Note that all the coordinate conversions occur 'under the hood' – the user inputs coordinates in the inertial frame, and will always receive output in the inertial frame as well.

(ii) At the beginning of each time-step, we evaluate $C$ and all $\mathcal{K}_{ij}$.

(a) If $C = 0$, this means no particles are currently undergoing pericenter passage. We use $M_1$.

(b) If $C = 1$, there is a particle undergoing a pericenter passage.

(1) If PARTIAL PERI is being used, use map $M_2$.

(2) If FULL PERI is being used, first convert back to inertial coordinates, then use map $M_3$. We are not performing any splitting here, so there is no need to do a post-time-step check. We always accept the step.

(iii) After executing $M_1$ or $M_2$, the conditions $C$ and $\mathcal{K}_{ij}$ are re-evaluated for each particle.

(a) If no particle pair that initially has $\mathcal{K}_{ij} = 0$ becomes $\mathcal{K}_{ij} = 1$, **and** $C$ does not go from 0 to 1, this means no close encounters of any sort have been introduced in the new step. We **accept** this step. Note that we do not concern ourselves with a particle pair with $\mathcal{K}_{ij} = 1$ initially going to $\mathcal{K}_{ij} = 0$ or $C$ going from 1 to 0, as this corresponds to a particle leaving the close encounter regime.

(b) Otherwise, a particle has entered a close encounter of some sort in the previous step. We **reject** the step, reset the simulation and perform a new step where the pre-encounter step now takes into account the updated $\mathcal{K}_{ij}$'s and $C$. The logic for the operators in this new step follow the first bullet point, accounting for the new values of $\mathcal{K}_{ij}$ and $C$ in this new step.

(iv) If needed (in the case of $M_1$ or $M_2$ being accepted), we convert from DHC to Inertial coordinates.

Note that for a step with the efficient map $M_1$ to be accepted, it must satisfy $\mathcal{K}_{ij} = 0$ and $C = 0$ both before and after the step. This ensures the time-reversibility of our algorithm – integrating in either time direction will result in the same switching between maps. In practice, very few steps will need to be rejected, typically of order a few per cent or less. But as we will see in the following section, the rejection of these few spurious steps results in very good long-term error performance, and the fact that so few steps need to be redone is a worthwhile trade-off.

We note that this algorithm is only almost perfectly time reversible. This is due to inconsistent or ambiguous cases that our algorithm cannot detect. For more discussion on this topic, see Hernandez & Dehnen (2023). We also note that even if the algorithm itself were to be perfectly time-reversible, floating-point precision and secular drift from Bulirsch–Stoer also render the algorithm not exactly time-reversible.

## 4.2 Switching functions

In this section, we describe the switching functions $\mathcal{K}\left(Q_{ij}^{(x)}\right)$ and $C\left(Q^{(x)}\right)$. In principle, any userdefined switching function that does not depend on the sign of time (for instance, a dependence on $\boldsymbol{V}^n$ with odd $n$) may be used – we will focus on the two switching functions used in our tests that are responsible for our algorithm's computational efficiency and are included in TRACE by default.

### 4.2.1 Planet–planet close encounter condition

The default switching condition is designed to detect close encounters at the beginning, end and during a time-step in a time-reversible way, making use only of the positions and momenta available at each time-step. Our algorithm approximates particle motion mid-timestep as straight lines, so some close encounters may be missed – however, we have found it to be effective in the vast majority of cases. Explicitly, the switching function which detects a close encounter between particles $i$ and $j$, $\mathcal{K}_{ij}\left(Q_{ij}^{(x)}\right)$, can be written:

$$\mathcal{K}\left(Q_{ij}^{(x)}\right) = \begin{cases} 1 \text{ for } Q_{\min} < a_{\mathrm{H}} R_{\mathrm{crit}} \\ 0 \text{ otherwise.} \end{cases} \tag{26}$$

On the right-hand side, $a_{\mathrm{H}}$ is a constant that may be set by the user ($a_{\mathrm{H}} = 3$ by default) and $R_{\mathrm{crit}}$ is the maximum of a modified Hill radius criteria between the two bodies,

$$R_{\mathrm{crit}} = \max\left(Q_i \sqrt[3]{m_i/3m_0}, \, Q_j \sqrt[3]{m_j/3m_0}\right), \tag{27}$$

which is the Hill radius where heliocentric distance replaces the traditional semimajor axis. The logic behind using the modified Hill radius condition is due to unbound particles: the Hill radius only has meaning for Keplerian orbits, and thus will not appropriately flag a close encounter between a pair of unbound planets. While less physically meaningful than the Hill radius, our criterion achieves good results across the board, and is much better for systems where particles become unbound, and the traditional Hill radius definition fails.

$Q_{\min}$ represents the minimum distance between particles over the range $t \in \left[t_n - \frac{h}{2}, t_n + \frac{h}{2}\right]$. We use a similar first-order estimate in time to the one described in Hernandez & Dehnen (2024,[5] section 3.1.1) to calculate $Q_{\min}$. To ensure time-reversibility, we must check in the two time directions by switching the sign of velocity. The minimum will occur in the $d$ time direction, where $d = \pm 1$, and $d = 1$ is forwards time and $d = -1$ is backwards time. $d$ satisfies $d(\boldsymbol{Q}_{ij} \cdot \boldsymbol{V}_{ij}) < 0$. In the rare case $\boldsymbol{Q}_{ij} \cdot \boldsymbol{V}_{ij} = 0$, $Q_{\min}$ is given at present, with $Q_{\min} = Q_{ij}$. Otherwise, $Q_{\min}$ occurs at $t = t_n + dt_{\min}$, with $t_{\min} = -d(\boldsymbol{Q}_{ij} \cdot \boldsymbol{V}_{ij})/V_{ij}^2$ (so that $t_{\min} > 0$). Then, we have

$$Q_{\min} = \begin{cases} \sqrt{Q_{ij}^2 - (\boldsymbol{Q}_{ij} \cdot \boldsymbol{V}_{ij})^2/V_{ij}^2} & \text{for } t_{\min} < h/2, \text{ and} \\ \sqrt{Q_{ij}^2 + hd(\boldsymbol{Q}_{ij} \cdot \boldsymbol{V}_{ij}) + h^2 V_{ij}^2/4} & \text{for } t_{\min} \geq h/2. \end{cases} \tag{28}$$

There are four key differences between the switching functions in TRACE and MERCURIUS: first, our switching function is discrete, while the MERCURIUS switching function is smooth. Secondly, MERCURIUS uses the standard Hill radius definition while ours is modified. Thirdly, MERCURIUS includes a more robust encounter prediction algorithm. In the MERCURIUS algorithm the particles are assumed to undergo Keplerian trajectories, rather than the straight lines used in TRACE. Due to the structure of the MERCURIUS integrator, it is feasible to perform a Kepler step 'for free' computationally, but this is impossible for TRACE, so we resort to more simplified trajectories. **This is the only advantage MERCURIUS has over TRACE.** Finally, MERCURIUS calculates the switching radius at the beginning of the integration for each pair of particles, which then remains fixed for the duration of the simulation. This is necessary to maintain the symplectic nature of MERCURIUS, but has the unfortunate side result of the switching radius becoming less physically meaningful if planet's semimajor axis changes over the course of the integration. However, changing the switching function does not impact reversibility, so this is not an issue for TRACE. To our knowledge, this is the first switching function for a hybrid integrator which **can** depend on the current state of the system, a novel result which greatly improves the flexibility of TRACE.

### 4.2.2 Pericenter condition

The recent work of Pham, Rein & Spiegel (2024) introduced a new adaptive time-step criterion for the IAS15 integrator. We use their result to inform our default choice of the pericenter switching. We first define,

$$\tau_{\mathrm{PRS},i} \equiv \sqrt{\frac{2 Q_i^{(2)} Q_i^{(2)}}{Q_i^{(3)} Q_i^{(3)} + Q_i^{(2)} Q_i^{(4)}}}, \tag{29}$$

where $Q_i^{(j)}$ is the magnitude of the $j$th derivative of heliocentric position of the $i$th particle. These higher order derivatives are approximated using a finite differencing method, the details of which are enumerated in Everhart (1985); Rein & Spiegel (2015). The necessity of calculating the higher-order derivatives does introduce some additional computational overhead, but this is not too significant and we judged that the robustness of this condition justifies this additional overhead. Then, our switching condition is given,

$$C\left(Q^{(x)}\right) = \begin{cases} 1 \text{ for } h > \eta \cdot \min_{i>1}\left(\tau_{\mathrm{PRS},i}\right) \\ 0 \text{ otherwise.} \end{cases} \tag{30}$$

Where we have found that $\eta = 1$ gives good results in our testing, and is hence set as the default value. Note that this condition is a minimum over all non-central bodies in the system. Thus if anybody is flagged for pericenter approach, the entire simulation will be integrated with BS or IAS15.

A note about our switching criteria follows. TRACE is a second order method. Defining the exact trajectory as $z(t)$, the TRACE trajectory as $\bar{z}_t(t)$, and the initial conditions $z(0)$, TRACE's local error is,

$$\bar{z}_t(h) = z(h) + \mathcal{O}(h^3). \tag{31}$$

The switching functions in this section can only be considered approximately physical due to this fact. By contrast, the orbit for a higher order method (like IAS15) is,

$$\bar{z}_h(h) = z(h) + \mathcal{O}(h^{n+1}), \tag{32}$$

with $n$ a larger integer like 15. For such higher order methods, the switching functions are more physical, representing time and length scales mimicking the orbits more closely. Regardless, our switching criteria work well in all tested problems.

These are the default pericenter switching conditions used in TRACE, but we have implemented a few others as well. See Appendix B for details on these alternatives.

---

[5]Hernandez & Dehnen (2024) have a typo in their section 3.1.1. Rather than $r_{\min} = \sqrt{q^2 + 2h_i \boldsymbol{p} \cdot \boldsymbol{q} + p^2 + h_i^2}$, it should read $r_{\min} = \sqrt{q^2 + 2h_i \boldsymbol{p} \cdot \boldsymbol{q} + p^2 h_i^2}$.

### 4.3 Collisions

Collisions and mergers constitute irreversible steps. Thus, TRACE cannot possibly be time reversible when collisions occur. TRACE handles collisions by enforcing a step acceptance: if a collision is detected mid-time-step, the step is automatically accepted regardless of either switching condition. TRACE is compatible with the standard REBOUND collision modules.

### 4.4 Additional forces

In many astrophysical systems it may be advantageous to consider extra dissipative forces which cannot be modelled through pure *N*-body gravitational interactions such as tides or radiation forces. It is very easy to add these additional forces to REBOUND using REBOUNDX (Tamayo et al. 2020), a library of additional physics that can be added to a REBOUND simulation.

We briefly comment on the efficacy of using a time-reversible integrator such as TRACE for the dynamics of the system if non-reversible forces are implemented. First, the extra forces do not necessarily need to be Hamiltonian—we can insert additional operators symmetrically in equation (10). These extra forces, as long as they are relatively small, do not compromise the long-term accuracy of a symplectic method as shown in Tamayo et al. (2020). Similarly, Hairer et al. (2006) show that time-reversible methods are not compromised by small non-Hamiltonian perturbations.

## 5 PERFORMANCE TESTS

In this section, we apply TRACE to a number of realistic astrophysical systems and compare its performance against other integrators available in REBOUND.[6] All the tests are performed using their C implementations. In all of our comparisons with IAS15, we use the new adaptive time-step criterion described by Pham et al. (2024). Unless otherwise specified, TRACE uses the FULL BS pericenter approach prescription for all our tests in this section.

### 5.1 Chaotic exchange orbit

We first investigate the case of a chaotic exchange orbit in the restricted coplanar three-body problem. The particular problem we have chosen includes a Sun-like star, a Jupiter-like planet on a circular orbit at its present-day semimajor axis, and a zero mass test particle. It has been studied in depth by a number of works including Wisdom (2017), Dehnen & Hernandez (2017), Hernandez (2019) and Hernandez & Dehnen (2023). In the circular restricted three-body problem, the only conserved quantity is the Jacobi constant $C_J$ (Murray & Dermott 2000; Tremaine 2023). For the initial value of $C_J$ we have selected in our tests, the test particle's orbit is exchanged between the primary and the secondary, undergoing multiple close encounters with the secondary. It can also never escape the system, making this problem an excellent test of body-body close encounters. Fig. 2 shows the results of our test. We integrate the system for 5000 orbits of the secondary, using a time-step of $h = 8$ days. We use a hill radius switching criteria of $a_H = 4.84$.[7] Once every 10 yr, the Jacobi constant error is calculated. We compare TRACE to MERCURIUS and WHFAST. As expected, WHFAST fails to resolve the close encounters with the secondary at all, and the error is immediately catastrophic.

[7]This value is selected for a direct comparison with MERCURIUS, which includes a hidden factor of 1.21 in the source code.

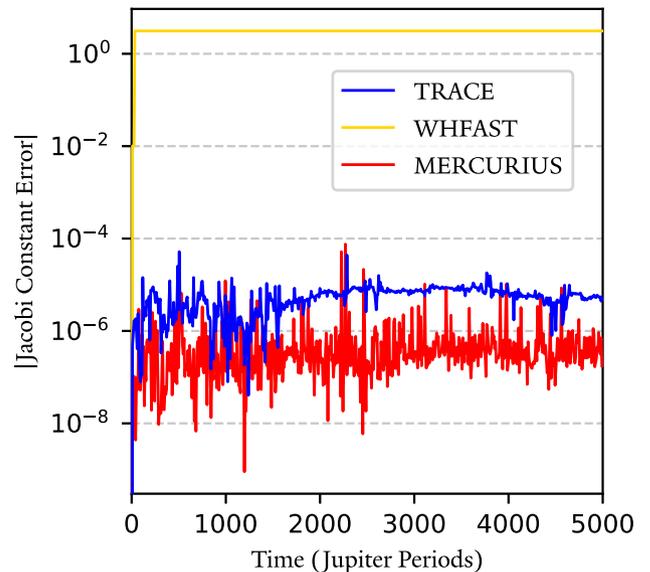

**Figure 2.** A comparison between TRACE, MERCURIUS, and WHFAST for the chaotic exchange problem. With a time-step of $h = 8$ d, the system is integrated for 5000 orbits of the primary with Jacobi constant error recorded every 100 yr. WHFAST immediately fails, while TRACE and MERCURIUS both show comparable good error performance with no secular drift. TRACE represents a 9.4x speed-up over MERCURIUS for this problem. The mean error for TRACE and MERCURIUS are $5.72 \times 10^{-6}$ and $3.01 \times 10^{-6}$, respectively.

MERCURIUS and TRACE are both able to resolve the close encounters, and both display very good error performance over the entire integration with no secular drift. The performance of both integrators are comparable, staying well below 1 per cent for the duration of the simulation. Comparing the runtimes of the two hybrid integrators: TRACE took 3.26 seconds while MERCURIUS took 39.7 s, a 12× speed-up.

### 5.2 Highly eccentric orbits

To evaluate TRACE's capabilities in resolving close encounters with the central body, we consider a two-planet system consisting of the Sun, Jupiter and Saturn. However, here Saturn's eccentricity is set to a various extremely high values, while its inclination is set to $\pi/2$ with respect to the orbit of Jupiter. With this setup, Saturn never has a close encounter with Jupiter, but does approach very close to the Sun.

This problem was first introduced in Levison & Duncan (2000) and revisited by Wisdom (2017); Hernandez & Dehnen (2023). We present the results of a number of tests we have performed on this system. First, we set $e = 0.99$ for Saturn and integrated the system for 300 Saturn orbits, using a time-step of $h = 0.15$ years. This is approximately 1/80th the period of Jupiter, the shortest orbital period of the system, so naively one might expect this to be an appropriate time-step for the Wisdom–Holman map. However, of course in reality this is not the case, as pericenter is not resolved. In the first panel of Fig. 3, we compare the performance of WHFAST, MERCURIUS, BS, and TRACE. The absolute value of the relative energy error, defined as $(E - E_{init})/E_{init}$, is output at the end of every time-step. We see that once again WHFAST immediately fails catastrophically, while MERCURIUS also fails to resolve the close encounters with the host star and the energy error rapidly exceeds $10^{-1}$. TRACE keeps the energy error well under $10^{-4}$ for the entire duration of the simulation, with no appreciable secular drift. BS does better than any of the splitting schemes, but secular drift is visible, which is expected of non–symplectic or non–reversible schemes. We also test WHFAST with a smaller time-steps which resolves the pericenter, to show that in principle it is possible to achieve similar results with the pure Wisdom–Holman map. The time-step necessary to resolve the pericenter for an eccentric orbit is related to the 'effective period at pericenter' (Wisdom 2015; Hernandez, Zeebe & Hadden 2022),

$$\tau_f = 2\pi \sqrt{\frac{(1-e)^3}{1+e} \frac{a^3}{Gm_0}}, \tag{33}$$

where $a$, $e$ are the eccentricity and semimajor axis of the orbit. We test WHFAST with a time-step of $\tau_f/50$. This time-step is chosen to achieve a close match with TRACE's performance. We refer to this tests as WHFAST Resolved. We see that it is possible to reach similar levels of energy error with pure Wisdom–Holman. However, we can see from the bottom panel of Fig. 3 that there are vast computational costs to picking such a small time-step, and that TRACE achieves similar error performance much faster.

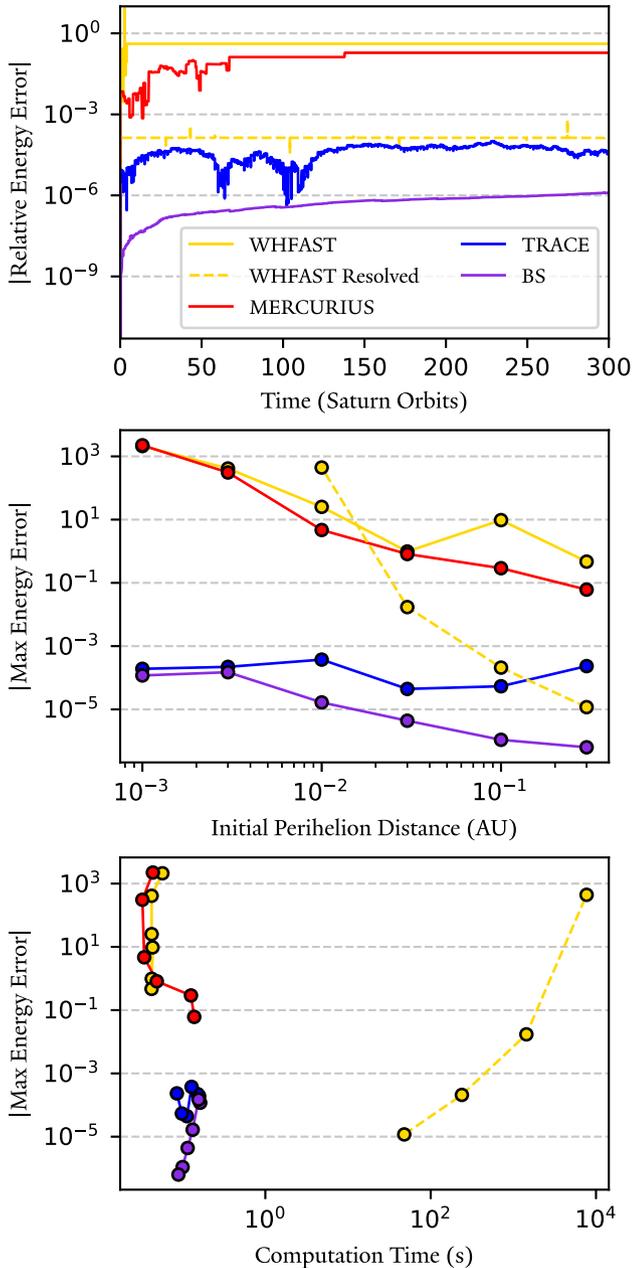

In the second panel of Fig. 3, we perform similar simulations, but we now set the initial eccentricity of Saturn to various values up to 0.9999. These values were selected for comparison with fig. 1 of Levison & Duncan (2000) and fig. 8 of Hernandez & Dehnen (2023). For each simulation, we plot the maximum energy error reached over the entire runtime. WHFAST and MERCURIUS perform similarly, with at least $10^{-1}$ energy error in all cases and reaching errors significantly greater than unity for the highly eccentric systems. TRACE again keeps the error around $10^{-3}$, and in fact shows consistent performance across all eccentricities. We see that even with much smaller time-steps WHFAST RESOLVED still performs badly at high eccentricities, while also having significantly slower compute time. The power of the pericenter switching condition allows TRACE to resolve extremely eccentric orbits with far more reasonable time-steps.

### 5.3 Violent systems

We envision violent systems to be one of the most relevant and powerful applications of TRACE. A violent system is one which undergoes significant dynamical instability, triggered by close encounters between planets. This can result in planets being ejected from the system, or being scattered onto orbits with high eccentricity and/or inclination. Planet-planet scattering almost certainly plays a role in sculpting the demographics of exoplanetary systems (Nagasawa & Ida 2011). Such systems are obvious applications for hybrid integrators, as for the vast majority of the simulation the planets are well separated and Wisdom–Holman is sufficient to accurately integrate the system. While close encounters in these systems represent a relatively small fraction of the total runtime, it is crucial to handle them with a conventional integrator to avoid catastrophic error. MERCURIUS is ineffective for many violent systems. While it can handle planet-planet close encounters well in most cases planets will be scattered onto highly eccentric orbits which leads to the pericenter approach not being resolved. In this section we will show that TRACE can handle these systems both quickly and accurately.

Let us consider a system of three Jupiter-mass planets orbiting a Sun-like star. Chambers, Wetherill & Boss (1996) showed that such a system will essentially always exhibit dynamical instability if their initial separations are less than 10 mutual Hill radii. To induce rapid dynamical instability in our system, we place the first planet at $a_1 = 5$ au and space the other two 3 mutual Hill radii out from the planet immediately interior. The eccentricity of each planet is set to 0.05, and the inclinations are set to $1°$, $2°$, $3°$ from the inner to outer planet. All other orbital angles are set to 0. We remove any particles which pass beyond $10^4$ au of the central star, using the `exit_max_distance` condition in REBOUND. Each time a particle is removed from the simulation, we reset to the center of mass frame of the system to avoid CoM and particle drift (which, left unchecked, would trigger the exit condition for all particles). We account for the lost energy associated with removing a particle from the system and the transformation back to the the new center of mass by using REBOUND's built in `track_energy_offset` feature. We consider collisions as well, using the built-in REBOUND collision modules `REB_COLLISION_DIRECT` for collision detection and `reb_collision_resolve_merge` for collision resolution. Collisions are flagged when any pair of particles overlap radii, and are resolved by merging the two colliding particles into one (conserving mass, momentum, and volume, but not energy).

**Figure 3.** A comparison of TRACE, MERCURIUS, BS and WHFAST for a Sun–Jupiter–Saturn system where Saturn has been given a high eccentricity. The first subplot shows the absolute value in the energy error over 300 orbits of Saturn, for a case where $e = 0.99$. MERCURIUS and WHFAST are both unable to resolve the close encounter with the Sun and the error quickly reaches or exceeds unity. TRACE keeps the error less than $10^{-3}$ with no secular drift. WHFAST RESOLVED, which are simply WHFAST with much smaller time-steps, is also plotted in dotted lines. It achieves similar error performance to TRACE, but requires significantly more time. The middle subplot compares the maximum error of six such systems with varying eccentricity for Saturn, plotted as a function of initial perihelion distance (the orbit slightly evolves over the course of the simulations). All MERCURIUS and WHFAST simulations perform worse as the eccentricity increases, while TRACE is consistently better. The bottom plot shows the same data points as the middle subplot, but now plotted on an efficiency cost versus maximum energy error plot. We see that while WHFAST RESOLVED can in principle keep up with TRACE in error performance for relatively low eccentricities, this comes at a vast computation cost. TRACE performs comparably to BS from a computation cost standpoint.

This is a highly chaotic system, so comparing the performance of integrators for a single system is essentially meaningless − the slight numerical differences ensure that we are very quickly working with entirely different systems. Instead, we take a statistical approach by considering an ensemble of such systems. The set-up of our analysis is as follows. We have run 500 instances of the system as described above, but have displaced the $x$-coordinate of the outermost planet by a random amount between $-10^{-12}$ and $10^{-12}$ au. We run each of these 500 instances with the following integrators: TRACE, MERCURIUS, BS, and IAS15. Each system is integrated forward in time for $10^7$ yr (which is roughly $9 \times 10^5$ initial orbits of the innermost planet). For TRACE and MERCURIUS we set the initial time-step equal to 0.221 yr. We arrive at this value from a conservation of energy argument. The smallest possible dynamical timescale at the end of this problem is the scenario where one planet is left on a close-in orbit and the other two are completely ejected. We calculate the orbital period of this close-in orbit and set our time-step to 1/15th of this value. IAS15 and BS are adaptive-time-step integrators − their initial time-steps are taken to be $2\pi \times 10^{-3}$ yr.

Fig. 4 shows the statistical results from our ensemble. The upper subplot displays the number planets in the system that survive over the course of the $10^7$ yr integration. The distributions of TRACE, BS, and IAS15 match each other very well, with the vast majority of systems ejecting one planet and retaining two. MERCURIUS, on the other hand, differs significantly in these statistics, with a more even split between one- and two-planet systems. The middle subplot displays the distribution of energy error at the end of the simulations of TRACE, BS, and MERCURIUS. As expected IAS15 performs very well, with a distribution centred around $10^{-11}$, and as such is omitted from the plot for clarity. MERCURIUS, also as expected, performs very poorly, with a median error very close to unity. Pure BS in general performs better than either hybrid integrator, with a median error of approximately $10^{-4}$. TRACE represents a significant improvement over MERCURIUS, with a median error of $10^{-1.49}$ compared to $10^{-0.14}$. The largest TRACE error is $10^{-0.28}$, and the largest MERCURIUS error is $10^{0.44}$. In the bottom subplot, we show histograms of the total runtime of simulations. For clarity, we only show the lower end of the IAS15 results − this distribution is centred on 26 min. The TRACE has a significant speed advantage over both BS and IAS15. This advantage grows the more particles are added to the system, as can be seen in the next section.

Of particular interest to those seeking to use TRACE on large ensembles of chaotic systems is the question of how well TRACE is able to reproduce the demographics of orbital elements on a statistical level. We investigate this in Fig. 5. For this phase, we consider only our simulations where two planets survive, as the other three cases do not have sufficient representation to perform robust statistical analysis on. We have plotted the cumulative distributions of eccentricity and inclination for the inner (P1) and outer (P2) planets for all simulations in which two planets survive, for each integrator. By eye, TRACE, BS, and IAS15 appear quite similar.

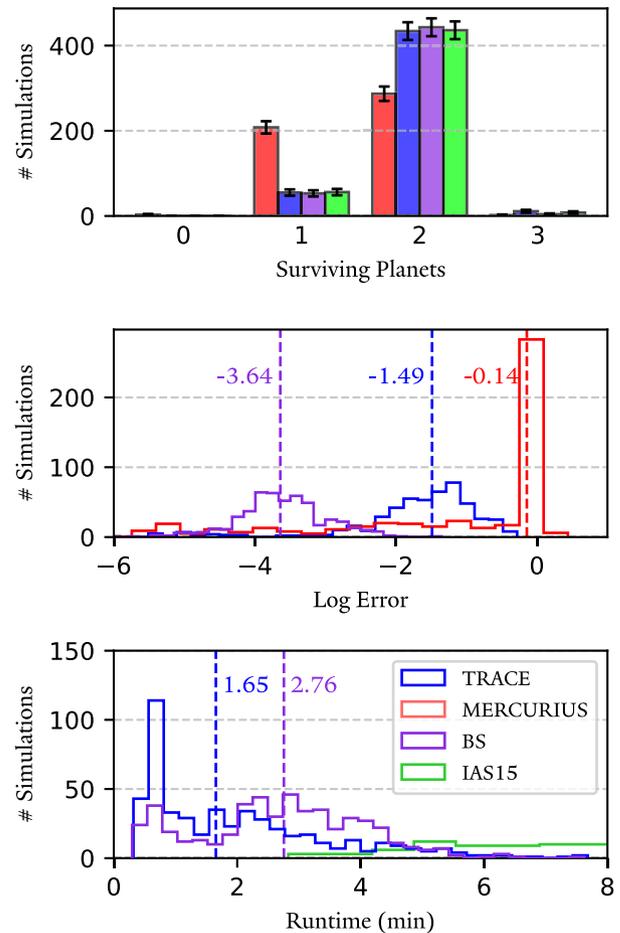

**Figure 4.** Statistics on an ensemble of 500 three-body scattering simulations, comparing the performance of TRACE, MERCURIUS, IAS15, and BS. The top subplot shows the final number of planets surviving at the end of the $10^7$ year integration. MERCURIUS is the obvious outlier, while TRACE replicates the statistics of IAS15 best. Poisson error bars are shown. The middle subplot shows histograms of the final energy errors, as well as the median values. IAS15 is not shown for clarity, but its distribution is centred around $10^{-11}$. TRACE represents a significant improvement over MERCURIUS, with the largest TRACE error being $10^{-0.28}$. The bottom subplot shows histograms of the runtimes. MERCURIUS is very fast due to the number of systems that eject too many planets. TRACE and BS both significantly improve on the IAS15 runtime, with TRACE having a small advantage over BS. The median runtimes of TRACE and BS are plotted in dotted lines. TRACE has a 1.67× speed advantage over BS and a 15.56× speed advantage over IAS15. Since MERCURIUS performs very poorly in this system we elect not to display its runtime on the bottom subplot to reduce clutter − it is very fast, but due to poor error performance should not be used for such systems.

In summary, TRACE reproduces the results of IAS15 quite well on a statistical level, with an almost 20× speed-up. This is in stark contrast to MERCURIUS, which qualitatively fails to reproduce the IAS15 statistics. In the preparation of this manuscript, Lu et al. (2024) used TRACE for a similar ensemble of violent scattering systems, with excellent error performance as well.

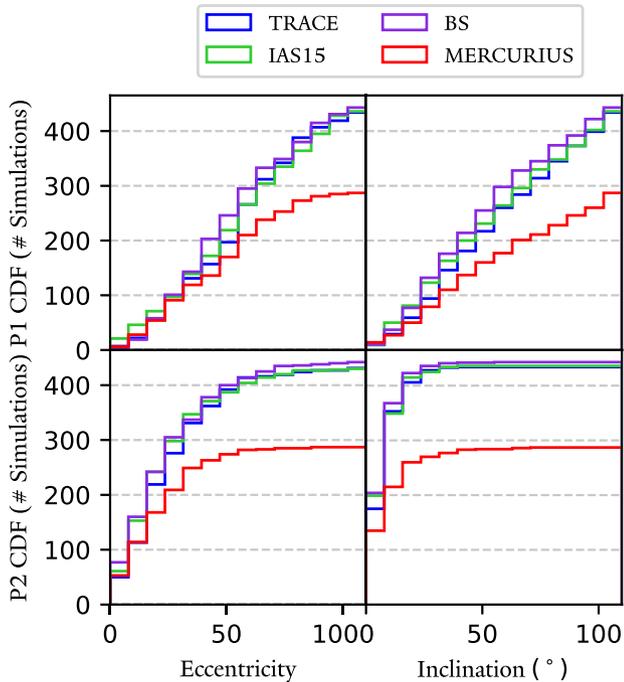

**Figure 5.** Cumulative distribution functions for the two-planet systems with all four integrators. The top row are the distributions for the inner planet, and the bottom row the distributions for the outer planet. We plot eccentricities in the left column and inclinations in the right. TRACE, BS, and IAS15 show very similar distributions, while it is clear that MERCURIUS underpredicts two-planet systems.

### 5.4 Accretion of the moon

As a test of TRACE's ability to integrate systems with a very large number of particles, we study the accretion of the moon from an impact disc generated by a giant impact via direct *N*-body simulations. This problem has been studied by Ida, Canup & Stewart (1997), Duncan et al. (1998), and Kokubo, Ida & Makino (2000) among others.

We present results from a simulation in the spirit of these studies. While our exact initial conditions do not match these studies, the final results are not sensitive to the precise initial conditions. Our simulation includes $10^3$ disc particles around an Earth-mass planet. The units of this simulation are the same as the study of Duncan et al. (1998): mass in Earth masses, Roche radius, and $G = 1$ (so a particle exactly at Earth's Roche radius has an orbital period of $2\pi$). The initial masses are randomly drawn from a power law distribution $\propto m^{-1}$ between $m = 3.2 \times 10^{-7}$ and $m = 3.2 \times 10^{-4}$. The total initial mass of the disc in our simulation is approximately four lunar masses. As in Kokubo et al. (2000), the density of the disc particles is $\rho_p = 3.3$ g cm$^{-3}$, while the density of the Earth is taken as $\rho_E = 5.5$ g cm$^{-3}$. Therefore, the radius of each disk particle is given by $r = (m/M_E)^{1/3}(\rho_p/\rho_E)^{-1/3}R_E$. In these units, $R_E = 1/2.9$. The semimajor axes of the disk particles are drawn from a power law distribution $\propto a^{-1}$ between $a = R_E$ and $a = 1.5$. The eccentricities, inclinations, and other orbital angles $(\Omega, \omega, f)$ are drawn from uniform distributions between $\{0, 0.95\}$, $\{0^\circ, 50^\circ\}$ and $\{0^\circ, 360^\circ\}$, respectively. Unlike Duncan et al. (1998), we do not remove initially Earth-crossing orbits as our method can handle highly eccentric orbits well.

We first do not consider collisions, and simply integrate the system with TRACE, MERCURIUS, IAS15, and BS for $6\pi$ time units. In this particular system while disk particles do get very eccentric, they have such low mass that error associated with a large jump term may not be significant. We thus also investigate the performance of TRACE using three prescriptions: the default FULL BS pericenter prescription, PARTIALPERI, as well as completely turning off pericenter switching. The results are plotted in Fig. 6 on an efficiency diagram. We see that TRACE, BS, and IAS15 all have very good error performance and high compute times. The poor computational performance of TRACE makes sense in this context: with so many particles we approach the limit of there being a pericenter close encounter every time-step – so TRACE essentially becomes BS with more overhead in this case. MERCURIUS is just as slow as the other three integrators, but performs worse due to failing to resolve pericenter approaches. Note, however, that despite failing to resolve pericenter approaches the error is still relatively good (around $10^{-3}$), as stated earlier. The reason for the slightly better performance of MERCURIUS compared to the non-pericenter TRACE prescriptions is due to MERCURIUS having a more robust encounter prediction routine, as described in Section 4. TRACE with no pericenter switching detected 77 299 close encounters, while MERCURIUS detected 137520. TRACE Partial and TRACE with no pericenter switching achieve similar accuracy to MERCURIUS, but significantly faster. TRACE Partial has a 6.2× speed advantage over MERCURIUS, and a TRACE with full pericenter switching at all has a 41.1× advantage. We conclude that for large *N* systems where the particles are relatively small, TRACE with full pericenter switching offers no advantage over BS or IAS15. But if we relax pericenter switching requirements, TRACE offers comparable (relatively good) error performance to MERCURIUS with a vastly improved runtime.

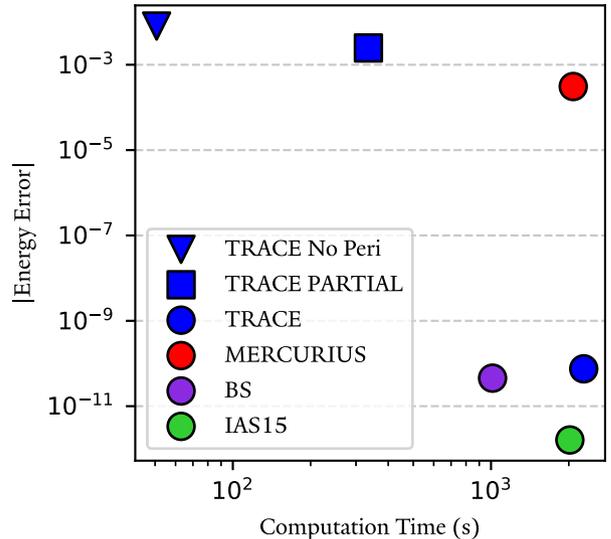

**Figure 6.** Efficiency plot of various integrators for the large-*N* accretion problem without considering collisions. TRACE, BS, and IAS15 all perform very well from an error standpoint, but are quite slow. MERCURIUS is just as slow, but is far less accurate. TRACE Partial and TRACE no Peri offer much faster alternatives while maintaining roughly the accuracy of MERCURIUS.

We now perform the same simulation, but integrate $10^3$ time units and turn collisions on with the same prescription as Section 5.3. Fig. 7 shows the results of our simulation using TRACE Partial. The left hand subplot shows the number of particles in the simulation as a function of time. While a direct comparison should not be made with the results of Ida et al. (1997) and Duncan et al. (1998) due to the slightly different initial conditions, qualitatively all four integrators match their results (and each other) well – see fig. 9 in Duncan et al. ([1998](#)). The right hand plots show snapshots of our TRACE simulation (blue) and IAS15 simulation (green) at the simulation's start (shared between the two simulations, plotted in black), 60 time units and at the end. The location of the particles think is plotted in cylindrical coordinates $(r, z)$ centred on Earth in units of Roche radius. These may be compared to figs 2–4 in Ida et al. ([1997](#)), and again are qualitatively similar. Both final results for TRACE and IAS15 in the bottom right panel shows one large body just within the Roche radii. This is in good agreement with the results of Ida et al. ([1997](#)) and Kokubo et al. ([2000](#)). The differences in our simulations can be attributed to differences in the initial conditions and integration methodology. The runtimes for the simulations are 7.06, 46.87, 90.13, and 65.98 seconds for TRACE, MERCURIUS, BS, and IAS15, respectively. TRACE improves on the runtime of MERCURIUS, BS, and IAS15 by 6.60×, 12.80x, and 9.30×, respectively.

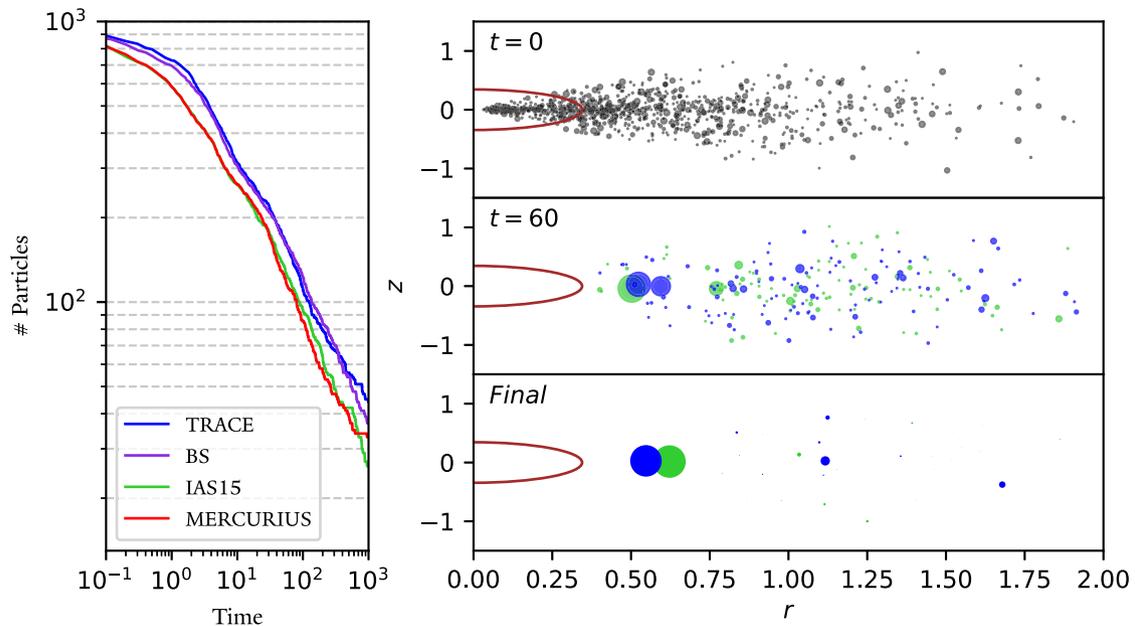

**Figure 7.** Results from a lunar accretion problem involving 1000 particles, accounting for collisions. We compare results from TRACE, MERCURIUS, IAS15, and BS. The left subplot shows the total number of bodies in the simulation as a function of time. All three integrators show good agreement. The right subplots show snapshots of the simulation at $t = 0$, 60 time units and the end of the simulation, for both TRACE (blue) and IAS15 (green). The positions of the planetesimals are plotted in cylindrical coordinates, and the size of each point in the graph corresponds to the planetesimal's mass. The physical radius of Earth is depicted in brown. We see that a roughly lunar-mass object (∼0.9 lunar masses in both simulations) forms at just within the Earth's Roche limit. Note also that many disc particles are initialized within the physical radius of Earth. In many previous works these particles would need to be removed due to poor resolution of pericenter passage – a significant advantage of TRACE is in the accurate handling of such particles.

We conclude that once again TRACE offers enormous computational benefits while maintaining acceptable levels of accuracy for large $N$ systems with collisions, qualitatively reproducing the results of IAS15.

### 5.5 ZLK cycles

The von Zeipel–Lidov–Kozai (ZLK) effect has been a well-studied phenomenon of great interest and wide application since its discovery (von Zeipel [1910](#); Kozai [1962](#); Lidov [1962](#); Naoz [2016](#)). In a hierarchical three-body system, a highly inclined outer perturber can induce significant coupled eccentricity and inclination oscillations in the orbit of the inner body. Similarly to Section [5.2](#) Wisdom–

Holman methods are in principle capable of accurately integrating the system. However, a worst-case time-step that accurately resolves the pericenter passage during high-eccentricity epochs must be applied over the length of the simulation, meaning that in practice it is actually faster to use adaptive-time-step higher order integrators such as TRACE, but we will show in this section that other integrators perform better.

We first consider a prototypical system in which ZLK oscillations are expected to occur. The initial values of our fiducial system are slightly modified from fig. 16 of Naoz ([2016](#)). In our test, we consider a Neptune-mass planet initially orbiting a 0.32 $M_\odot$ star with $a_1 = 2$ au and $e_1 = 0.01$. The perturber is a 10 $M_J$ brown dwarf orbiting the primary with $a_2 = 50$ au, $e_2 = 0.52$ and $i_2 = 80°$. We integrate this system with IAS15, WHFAST, and TRACE. For TRACE, we use a time-step equal to $1/20$ the initial orbital period of the planet. Fig. [8](#) plots the eccentricity evolution of the inner planet and the energy error over two ZLK cycles.

The runtimes for TRACE, WHFAST, and IAS15 are 19.63, 7.47, and 284.98 s, respectively. We see that while TRACE maintains an acceptable level of error in this problem, WHFAST actually outperforms it in both speed and computation time. The reason for this has to do with the choice of splitting scheme and coordinates. By default, WHFAST is implemented in Jacobi coordinates. This differs from DHC used by TRACE in that $\mathcal{H}_K$ does exactly represent a Keplerian orbit. Hence, in Jacobi coordinates Wisdom–Holman methods are able to accurately integrate arbitrarily eccentric orbits without the need to choose an extremely small time-step to resolve pericenter approach. We demonstrate this by also plotting an implementation of WHFAST in DHC in dotted lines, with the same time-step. We see that this scheme fails at relatively low eccentricity, and the planet is quickly ejected from the system.

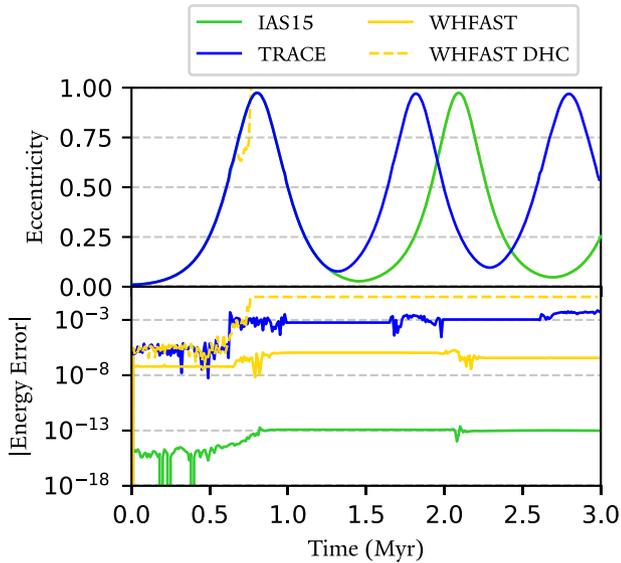

**Figure 8.** A comparison of IAS15 (green), TRACE (blue), and WHFAST in both Jacobi (gold) and DH (gold dashed) coordinates for ZLK oscillations. The upper subplot shows eccentricity evolution, and the bottom subplot depicts energy error over time. The eccentricity evolution of IAS15 and WHFAST in Jacobi coordinates are identical at this scale – their curves appear identical. WHFAST in Jacobi coordinates outperforms TRACE in both error and computation performance. TRACE gives qualitatively different results than IAS15 and WHFAST.

We emphasize that this is not exactly a failure case for TRACE, as it is working as intended. Rather, this should be seen as a strength of the Wisdom–Holman method for the specific case of a highly eccentric innermost planet. We conclude that for such systems, TRACE is unsuitable: if one desires extremely high accuracy IAS15 should be used, and if moderate accuracy with high speed is required then WHFAST outperforms TRACE and should be used instead.

## 6 POTENTIAL IMPROVEMENTS

In this section, we list two potential improvements to TRACE that are not currently implemented.

### 6.1 Pairwise reversibility

Currently, upon rejection of a time-step, TRACE will reset the entire simulation to the initial state and re-integrate all particles. This is unavoidable for close encounters with the central body, since the TRACE map entails moving the entire $\mathcal{H}_J$ to $\mathcal{H}_K$, which would couple the equations of motions of all the particles and necessitates solving them all with BS. However, this in principle can be avoided for planet–planet close encounters, since moving $\mathcal{H}_I$ to $\mathcal{H}_K$ is only a function of the positions and velocities of the two bodies undergoing a close encounter. Hence, we should only need to redo the interaction steps for the non-close encounter particles, without needing to recalculate the relatively expensive Kepler step.

In practice this is only a time save for step rejections, which for the majority of simulations are a comparatively small fraction of the total steps taken in the simulation, so the actual computational benefit is insignificant. We have hence elected to not include pairwise reversibility in this iteration of TRACE.

### 6.2 Adaptive time-stepping

Hernandez & Dehnen (2023) showed that reversibly switching between time-steps using the same switching scheme is feasible. Notably, Hernandez & Dehnen (2024) were able to reversibly adapt the time-step of a SYMBA-like algorithm to great effect. Their implementation used different time-steps for different 'shells' of increasing distance from the host star. The difficulty of a more flexible scheme valid for a wider array of astrophysical systems precluded its inclusion into TRACE. In principle, the global time-step of TRACE should be able to be adapted reversibly, which would result in performance gains.

## 7 CONCLUSION

We present TRACE, a time-reversible hybrid integrator capable of efficiently and accurately resolving any type of close encounter in the planetary $N$-body problem. TRACE matches or improves upon the error performance of current hybrid integrators such as MERCURIUS with a conceptually simpler switching scheme and a significant speedup (up to 14× for certain problems). TRACE is freely available as part of the REBOUND package at https://github.com/hannorein/rebound.[8] It is available in both C and PYTHON. From our testing, TRACE is superior to MERCURIUS in all cases. We anticipate TRACE having a myriad of useful applications including violent scattering systems, large $N$ systems, and systems with highly eccentric orbits.

While TRACE shows excellent performance, there are clear avenues of improvement such as pairwise-reversibility and adapting the global time-step, both of which could lead to significant speedups. The fact that TRACE is almost completely reversible lends itself to significantly more flexibility than symplectic integrators such as MERCURIUS, in particular with our choices of switching functions. We did not deeply explore potential switching functions – rather, we aimed to select safe defaults for the user. In principle, these switching functions could be any arbitrary function of particle positions and velocities and further exploration may lead to better results.

It is instructive to directly compare and discuss the advantages TRACE has over MERCURIUS, the current hybrid integrator implemented in REBOUND. First, TRACE is significantly faster than MERCURIUS in many cases. There is some minor benefit from the simpler switching function. However, the vast majority of the speedup comes from the use of BS for close encounters in the case of TRACE, instead of IAS15 for MERCURIUS. While IAS15 indeed is significantly more accurate than BS, for hybrid integrators the error is dominated by error associated with operator splitting (see Section 2.1). This greatly overshadows the difference in error between IAS15 and BS, so our choice of BS over IAS15 provides significant speed benefits with negligible accuracy tradeoff. Secondly, TRACE is able to resolve close encounters with the central body, unlike MERCURIUS which can only handle close encounters between pairs of planets. This allows TRACE to effectively integrate highly eccentric orbits which MERCURIUS fails at. Finally, ignoring finite floating point precision, TRACE is exactly time-reversible while MERCURIUS is symplectic. The almost time-reversible nature of TRACE means that it has good long-term error conservation properties as we have shown in this work. The fact that TRACE is not symplectic affords it several flexibility advantages over MERCURIUS, which is demonstrated most prominently in the switching function. To maintain the symplectic nature of MERCURIUS, the switching distance is set at the beginning of the simulation and cannot change. If the system significantly changes this criteria may become unphysical – for instance, if a planet moves outward over the course of the simulation close encounters will be underpredicted. TRACE does not face this issue, and can adjust the switching criteria as a function of the state of the simulation such that it always remains a physically meaningful quantity. This has the further advantage that we can implement TRACE such that each time-step only depends on the inertial particle coordinates, but not pre-calculated per-particle parameters such as switching radii. This makes adding/removing/colliding/merging particles during a simulation much easier. Given that TRACE performs strictly better than MERCURIUS, the MERCURIUS integrator will be depreciated in the near future.

[8]Extensive documentation and example notebooks are available at https://rebound.readthedocs.io.

Finally, we discuss specific use cases for TRACE. We must emphasize that by virtue of being a hybrid integrator TRACE has limited use cases. In the vast majority of cases, REBOUND users are encouraged to use WHFAST for the planetary *N*-body problem when there are no close encounters, or IAS15 for a wider variety of problems where high accuracy is paramount. We recommend TRACE for cases of the planetary *N*-body problem where close encounters do occur, be it with the central star or between pairs of planets, and where only moderate accuracy is required but fast computation is desired. For instance, in large ensembles of chaotic systems exact accuracy in each individual system is not required to recover macroscopic quantities on a statistical or population level – and very large ensembles benefit greatly from the speed-ups afforded by TRACE.

## ACKNOWLEDGEMENTS

We thank the reviewer Antoine Petit for helpful comments that greatly improved the manuscript. We thank Yasmeen Asali, Walter Dehnen, Shigeru Ida, Gregory Laughlin, Rosemary Mardling and J. Sebastian Monzon for enlightening discussions. This work has benefited from use of the GRACE computing cluster at the Yale Center for Research Computing (YCRC).

## DATA AVAILABILITY

The TRACE integrator is publicly available as part of the REBOUND package, which can be installed at https://github.com/hannorein/rebound. Extensive documentation is available at https://rebound.readthedocs.io. The code used to make the figures in this paper can be found at https://github.com/tigerchenlu98/rebound/tree/TRACE/examples/TRACE_paper.

## APPENDIX A: DEMOCRATIC HELIOCENTRIC COORDINATES

`TRACE` uses the Democratic Heliocentric Coordinates introduced by Duncan et al. (1998). These coordinates are also used by `MERCURIUS`, and are given by

$$Q_i = \begin{cases} q_i - q_0 & \text{for } i \neq 0 \\ \frac{1}{m_{tot}} \sum_{j=0}^{N-1} m_j q_j & \text{for } i = 0. \end{cases} \quad (A1)$$

The corresponding conjugate momenta are given

$$P_i = \begin{cases} p_i - \frac{m_i}{m_{tot}} \sum_{j=0}^{N-1} p_j & \text{for } i \neq 0 \\ \sum_{j=0}^{N-1} p_j & \text{for } i = 0. \end{cases} \quad (A2)$$

## APPENDIX B: OTHER PERICENTER SWITCHING FUNCTIONS

In this Appendix, we describe some alternative prescriptions for our pericenter switching condition that are also included with `TRACE`.

### B0.1 Effective period at pericenter

Wisdom (2015) demonstrated that the Wisdom–Holman method is able to integrate arbitrarily eccentric orbits, so long as the time-step chosen does not exceed 1/16th of the effective period at pericenter, or $2\pi/\dot{f}|_{\text{pericenter}}$ where $f$ is the true anomaly. We use this result to inform another possible choice of pericenter switching condition:

$$P_i = \frac{2\pi}{\dot{f}_i} - a_p h, \quad (B1)$$

where

$$\dot{f}_i = \frac{|Q_i \times v_i|}{Q_i^2}. \quad (B2)$$

$a_p$ is a constant that may be set by the user. Wisdom (2015) recommends $a_p = 17$. The full switching condition is given by

$$C\left(Q^{(x)}\right) = \begin{cases} 1 \text{ for } \min_{i>0} P_i < 0 \\ 0 \text{ otherwise.} \end{cases} \quad (B3)$$

While powerful, this condition is incomplete – it is only meaningful for bound Keplerian orbits. For unbound orbits, this condition does not trigger.

### B0.2 Heliocentric distance

The most simple prescription one can use is simple heliocentric distance from the star, and can be written,

$$C\left(Q^{(x)}\right) = \begin{cases} 1 \text{ for } \min_{i>0}[Q_i < a_P] < 0 \\ 0 \text{ otherwise.} \end{cases} \quad (B4)$$

The choice of $a_P$ is not intuitive, depends on the scale of the system, and may require some experimentation. However, if a suitable value is found for a particular system, this condition offers the most easily understood pericenter switching condition.

### B0.3 None

It is also possible to turn off pericenter switching as a whole, which may be desirable for some problems. In this case, `TRACE` essentially becomes a faster version of `MERCURIUS`.

## APPENDIX C: SYMPLECTICITY OF TRACE

In this Appendix, we demonstrate that although `TRACE` is composed of maps that are both symplectic and reversible, switching between them renders `TRACE` itself not exactly symplectic. Assume we have symplectic and reversible maps $M_1$ and $M_2$ and we'd like to switch between them. $z_i$ is defined as the mapped phase space and $z$ is the initial state: $z_i = M_i z$. By definition of symplecticity, it follows that,

$$\Omega = J_i^T \Omega J_i, \quad (C1)$$

where $\Omega$ is an antisymmetric matrix,

$$\Omega = \begin{bmatrix} 0 & -1 \\ 1 & 0 \end{bmatrix}, \quad (C2)$$

and $J_i$ is the Jacobian matrix, $J_i = \partial z_i / \partial z$ for $i = 1$ and 2. Let's switch according to some binary scalar function $F(z)$, taking values 0 or 1, in a way that is non-reversible. We merely want to prove that switching between two symplectic maps is not symplectic, without concern for reversibility. Denote a map that switches between $M_1$ and $M_2$ as $M_3$:

$$z_3 = M_3 z = (M_1 F(z) + M_2 (1 - F(z))) z, \quad (C3)$$

Note that this map behaves differently from the algorithm behind the `TRACE` code; in the code's case, $F$ is set before and after a global time-step (a modified Heaviside function), but here $F$ can vary within a step. We return to this difference shortly. The Jacobian of this map is

$$J_3 = \frac{\partial z_3}{\partial z} = J_1 F(z) + J_2 (1 - F(z)) + (z_1 - z_2) \frac{\partial F}{\partial z}. \quad (C4)$$

$\partial F / \partial z$ is not finite everywhere and thus $M_3$ is not canonical (symplectic) everywhere. Because a modified Heaviside map has similar discontinuous $F$ behaviour, it follows that the algorithm behind the `TRACE` code is also non-symplectic. Given what we have just shown, we can argue that we can switch between two symplectic methods in a nearly symplectic way, but experimentally such an algorithm often shows poor error performance. Establishing near time-reversibility, as `TRACE` does, is crucial for good error performance in switching between the two symplectic maps discretely.

### C1 Symplecticity of variable time-step maps

In fact, we can also show that symplectic maps with variable time-steps also lose exact symplecticity. We illustrate this as follows. Write

a general map as

$$q_1 = a(z, h),$$
$$p_1 = b(z, h),$$

(C5)

where we assume for the time being that $h$ is a constant parameter. We can form the Jacobian of (C5), and insert in the condition (C1). The constraint so that map (C5) is symplectic is

$$\partial_q a \partial_p b - \partial_q b \partial_p a = 1,$$

(C6)

which is simply the requirement that the determinant of the Jacobian matrix be 1. Now, assume $h$ is no longer a constant parameter; e.g.

$$q_1 = a(z, h(z))$$
$$p_1 = b(z, h(z)).$$

(C7)

We take the Jacobian of map (C7), again insert into (C1), and use the new constraint (C6). We are left with the following constraint, assuming $a \neq b$:

$$\partial_q h \partial_p b \partial_p a - \partial_p h \partial_q a \partial_q b = 0.$$

(C8)

Only special $h(z)$ satisfying equation (C8), depending on the specific map functions $a$ and $b$, constitute symplectic maps. It would be interesting to construct such functions to test this constraint.

This paper has been typeset from a TeX/LaTeX file prepared by the author.